\documentclass[11pt,a4paper]{article}
\usepackage[T1]{fontenc}
\usepackage[utf8]{inputenc}
\usepackage{lmodern}
\usepackage[margin=25mm]{geometry}
\usepackage{amsmath,amssymb}
\usepackage{graphicx}
\usepackage{xcolor}
\usepackage{authblk}
\usepackage[numbers,sort&compress]{natbib}
\usepackage{caption}
\usepackage[colorlinks=true,allcolors=blue]{hyperref}
\usepackage{tikz}
\usetikzlibrary{arrows.meta,positioning,fit,calc,shapes.geometric}
\usepackage{tabularx}
\usepackage{array}
\usepackage{adjustbox}
\usepackage{array}
\usepackage{booktabs}
\usepackage{tabularx}

\newenvironment{keywords}{\par\smallskip\noindent\textbf{Keywords: }}{\par\medskip}
\newenvironment{acknowledgements}{\section*{Acknowledgements}}{}

\tikzset{
  dtbox/.style={draw, rounded corners=2pt, align=center,
    minimum height=8mm, text width=25mm, inner sep=3pt},
  dtwide/.style={dtbox, text width=34mm},
  dtarrow/.style={-{Latex[length=2.2mm]}, line width=0.7pt},
  dtfeedback/.style={dtarrow, dashed},
  dtgroup/.style={draw, rounded corners=3pt, inner sep=5pt}
}

\begin{document}

\title{\textit{Mus siliconus}: A Neuro-Musculoskeletal Digital Twin of the
Mouse Integrating Neural Dynamics, Biomechanics, and Tactile Sensing}
\hypersetup{
  pdftitle={Mus siliconus: A Neuro-Musculoskeletal Digital Twin of the Mouse Integrating Neural Dynamics, Biomechanics, and Tactile Sensing},
  pdfauthor={Satoshi Oota, Hideo Yokota, Hiroki Mori},
  pdfsubject={Position paper on an embodied neuro-musculoskeletal digital twin of the mouse},
  pdfkeywords={digital twin, neuro-musculoskeletal modeling, computational neuroscience, tactile sensing, neurorobotics}
}
\author[1]{Satoshi Oota\thanks{Corresponding author: \href{mailto:oota@riken.jp}{oota@riken.jp}}}
\author[1]{Hideo Yokota}
\author[2]{Hiroki Mori}

\affil[1]
{Image Processing Research Team, Center for Advanced Photonics (RAP), RIKEN,
2-1 Hirosawa, Wako, Saitama 351-0198, Japan}
\affil[3]
{\textit{Affiliation to be confirmed}}
\affil[2]
{Future Robotics Organization, Waseda University, 3-4-1 Okubo, Shinjuku City, Tokyo 169-0072, Japan.}

\date{}

\maketitle

\begin{abstract}
Digital twin technologies offer the potential to transform neuroscience and biomedicine by creating predictive computational representations of living organisms. However, most existing animal digital twins focus on isolated components such as neural circuits, anatomy, or biomechanics, rather than the integrated processes that give rise to behavior. We argue that future animal digital twins should be conceived as embodied dynamical systems in which neural activity, body mechanics, sensory feedback, and environmental interactions are modeled within a unified framework.

To support this vision, we propose a neuro-musculoskeletal digital twin of the mouse that integrates multimodal anatomical reconstruction from X-ray CT, high-resolution white-light section data, and Scx-GFP imaging with biomechanical simulation, Bonhoeffer--van der Pol neural dynamics, and tactile sensory feedback. The resulting framework forms a closed-loop sensorimotor architecture in which behavior emerges from continuous interactions among the nervous system, musculoskeletal system, and environment. We further argue that the Bonhoeffer--van der Pol model provides a useful dynamical-systems foundation for describing these interactions while remaining computationally tractable for large-scale simulations.

Beyond its application as a modeling framework, we envision neuro-musculoskeletal digital twins as a convergence point for computational neuroscience, biomechanics, artificial intelligence, and robotics. Coupled with adaptive learning algorithms and autonomous experimental platforms, future digital twins may evolve from passive simulation environments into active scientific instruments capable of generating hypotheses, predicting interventions, and guiding experimental discovery. We propose that such embodied digital twins represent an important step toward understanding biological intelligence and developing the next generation of adaptive biomedical and robotic systems.
\end{abstract}


\begin{keywords}
Digital Twin |
Neuro-Musculoskeletal Modeling |
Bonhoeffer--van der Pol Model |
Embodied Intelligence |
Computational Neuroscience |
Biomechanics |
Tactile Sensing |
Reinforcement Learning |
Autonomous Robotics
\end{keywords}


\section*{Introduction}

The emergence of digital twin technologies is transforming the study of complex biological systems \cite{sun2022digital,drummond2024patient}. By combining computational modeling with experimental observations, digital twins offer the possibility of creating predictive representations of living organisms that can be continuously updated, interrogated, and validated \cite{sun2022digital,hood2004predictive}. In neuroscience and biomedicine, digital twins have been proposed as tools for understanding disease progression, predicting treatment outcomes, and integrating multimodal biological data \cite{drummond2024patient,garanin2025digital}. However, most existing efforts focus on isolated aspects of biological function, such as neural circuits, anatomy, or biomechanics, rather than the integrated processes that give rise to behavior \cite{appukuttan2023ebrains}.

The mouse is one of the most important model organisms in modern biology and neuroscience \cite{guenet2005mouse,rosenthal2007mouse}. Advances in multimodal imaging, neural recording technologies, behavioral tracking, and biomechanical modeling have created unprecedented opportunities for constructing comprehensive digital representations of mouse anatomy and physiology \cite{steinmetz2018neuropixels,pereira2019fast,delp2007opensim}. Nevertheless, current computational models remain fragmented. Neural models often neglect the body and environment, biomechanical simulations frequently rely on prescribed control strategies, and sensory systems are commonly treated as secondary components \cite{dayan2001theoretical,delp2007opensim,todorov2012mujoco}. As a result, many existing models fail to capture the continuous interactions among neural activity, body mechanics, sensation, and environmental feedback that characterize biological behavior \cite{beer2000dynamical,pfeifer2006body}.

We argue that future animal digital twins should be conceived as embodied dynamical systems rather than collections of independently modeled subsystems \cite{pfeifer2006body,beer2000dynamical}. From this perspective, behavior emerges from interactions among neural processes, musculoskeletal dynamics, sensory feedback, and environmental constraints \cite{beer2000dynamical,kelso1995dynamic}. Understanding these interactions requires a common mathematical framework capable of describing the coupled dynamics of the nervous system, body, and environment \cite{thelen1994dynamic,izhikevich2007dynamical}.

In this position paper, we propose the Bonhoeffer--van der Pol (BVP) model as a conceptual and computational foundation for next-generation neuro-musculoskeletal digital twins. The BVP model was introduced by FitzHugh as a reduced nonlinear model of excitable and oscillatory nerve-membrane dynamics and was subsequently implemented as an equivalent electrical circuit by Nagumo and colleagues \cite{fitzhugh1961impulses,nagumo1962active}. Its low-dimensional structure captures essential phenomena such as excitability, refractoriness, threshold behavior, and self-sustained oscillation without reproducing every biophysical detail of the Hodgkin--Huxley model \cite{hodgkin1952quantitative}.

Building on this abstraction, we propose that neural activity, motor coordination, tactile sensation, and adaptive behavior can be modeled as emergent phenomena arising from interactions among coupled nonlinear oscillators distributed throughout an embodied system. This proposal is consistent with evidence that coupled neural oscillators and central pattern generators can produce and coordinate rhythmic motor behavior in animals and robots \cite{ijspeert2008central}, as well as with dynamical-systems accounts in which cognition and behavior emerge from continuous interactions among neural, bodily, sensory, and environmental processes \cite{beer2000dynamical,buhrmann2013dynamical}. Such an embodied dynamical perspective shifts the emphasis from exhaustive representation of individual biological components toward identifying the organizational principles governing their interactions. Embedding this framework within a digital-twin architecture could therefore complement anatomically detailed neuromusculoskeletal models with compact dynamical models of neural control, sensory feedback, and behavioral adaptation \cite{saxby2023digital}.

Building upon this foundation, we propose a neuro-musculoskeletal mouse digital twin that integrates four key elements: (i) multimodal anatomical reconstruction from X-ray CT, high-resolution white-light serial-section data, and Scx-GFP imaging; (ii) biomechanical simulation of the musculoskeletal system; (iii) Bonhoeffer--van der Pol neural dynamics; and (iv) tactile sensory feedback arising from interactions with the environment. The feasibility of integrating CT and color cryosection data into a common three-dimensional mouse atlas has previously been demonstrated \cite{dogdas2007digimouse}, while Scx-GFP reporter mice provide fluorescence-based visualization of tendon and ligament structures \cite{pryce2007transgenic}. Existing whole-body mouse musculoskeletal models further provide a foundation for anatomically grounded biomechanical simulation and the integration of neural control, body dynamics, sensory feedback, and environmental forces \cite{ramalingasetty2021mouse}.

Together, these components form a closed-loop sensorimotor architecture in which neural activity generates movement, movement alters bodily and environmental states, and tactile signals feed information back to the controller. This formulation is consistent with established sensorimotor-control frameworks emphasizing the continuous coupling of sensing and action \cite{buhrmann2013dynamical,seminara2023hierarchical}, as well as with embodied approaches in which adaptive behavior emerges through interactions among the nervous system, body, and environment \cite{beer2000dynamical,pfeifer2006body}.

More broadly, we envision neuro-musculoskeletal digital twins as a bridge among computational neuroscience, biomechanics, artificial intelligence, and autonomous robotics. Digital-twin frameworks are increasingly being considered as integrative and predictive tools in neuroscience and neuromusculoskeletal research \cite{saxby2023digital,fekonja2024digital}. By combining anatomically grounded models with dynamical-systems theory, machine learning, and robotic experimentation, future digital twins may evolve from passive simulation environments into active scientific instruments capable of generating hypotheses, predicting the effects of interventions, and guiding experimental discovery. This vision is supported by developments in robot scientists and self-driving laboratories, which have demonstrated that artificial-intelligence systems can select hypotheses, design experiments, evaluate results, and iteratively determine subsequent experiments \cite{king2009automation,hase2019next}.

\section*{Conceptual Framework}

The central premise of this position paper is that animal behavior emerges from continuous interactions among the nervous system, the body, and the environment \cite{chiel1997brain,beer2000dynamical}. Neural circuits do not operate in isolation: their sensory inputs and motor outputs are shaped by musculoskeletal mechanics, the physical properties of the body, and interactions with the external world \cite{nishikawa2007neuromechanics,tytell2011spikes}. We therefore propose that future neuro-musculoskeletal digital twins should be constructed as embodied dynamical systems rather than as collections of independently modeled components. In such systems, neural activity, musculoskeletal mechanics, sensory processing, and environmental interactions are tightly coupled through closed-loop feedback mechanisms \cite{buhrmann2013dynamical}. This integrated formulation is consistent with the emerging concept of neuromechanical digital twins, in which artificial neural controllers are embedded within physically realistic body models interacting with simulated environments \cite{wangchen2026embodied}.

The proposed framework consists of four interconnected domains: neural dynamics, musculoskeletal mechanics, sensory feedback, and environmental interactions. Neural activity generates motor commands that activate muscles and produce movement, while the resulting behavior is constrained and shaped by the mechanical properties of muscles, the body, and the external environment \cite{nishikawa2007neuromechanics,tytell2011spikes}. Body motion, in turn, changes the animal's interactions with its surroundings, generating tactile and proprioceptive signals that influence subsequent neural activity and motor output \cite{proske2012proprioceptive,johansson2009tactile}. This recurrent interaction reflects the integration of feedforward neural control with sensory feedback and intrinsic body mechanics observed in animal locomotion \cite{ijspeert2023integration}. Behavior therefore arises not from any single subsystem, but from the continuous exchange of information and forces among the nervous system, body, and environment \cite{chiel1997brain}.

At a conceptual level, the framework can be summarized as a recurrent perception--action loop in which neural activity drives body mechanics, bodily action changes the environment, and the resulting sensory signals modify subsequent neural activity \cite{chiel1997brain,tytell2011spikes,buhrmann2013dynamical}:

\begin{equation}
\begin{gathered}
\text{Neural Dynamics}
\rightarrow
\text{Body Mechanics}
\rightarrow
\text{Environment}
\\
\rightarrow
\text{Sensory Feedback}
\rightarrow
\text{Neural Dynamics}
\end{gathered}
\label{eq:conceptual_loop}
\end{equation}

This closed-loop architecture provides a common foundation for integrating multimodal anatomical data, biomechanical simulation, Bonhoeffer--van der Pol neural dynamics, and tactile sensing within a single computational framework. Previous studies have separately demonstrated multimodal three-dimensional mouse reconstruction \cite{dogdas2007digimouse}, whole-body mouse musculoskeletal simulation incorporating neural commands and sensory feedback \cite{ramalingasetty2021mouse}, low-dimensional modeling of excitable neural dynamics \cite{fitzhugh1961impulses,nagumo1962active}, and the use of tactile information in closed-loop sensorimotor control \cite{johansson2009tactile,seminara2023hierarchical}. Our proposed framework brings these elements together as an embodied dynamical system.

Importantly, the objective is not merely to reproduce movement or neural activity in isolation, but to capture the coupled neural, mechanical, sensory, and environmental dynamics from which adaptive behavior emerges \cite{nishikawa2007neuromechanics,ijspeert2023integration}. Such an approach recognizes that neural activity alone is generally insufficient to explain behavior because its functional consequences depend on body mechanics, sensory feedback, and interactions with the physical environment \cite{tytell2011spikes}.

More broadly, we view this framework as a bridge among computational neuroscience, biomechanics, artificial intelligence, and robotics. This interdisciplinary perspective is consistent with neuromechanical approaches in which animal models and robots serve as complementary tools for investigating autonomous behavioral control \cite{ramdya2023neuromechanics}, as well as with digital-twin frameworks that integrate mechanistic simulation, subject-specific data, and predictive modeling \cite{saxby2023digital}. By enabling continuous interactions among perception, action, body mechanics, and environmental feedback, neuro-musculoskeletal digital twins may provide a foundation for studying embodied intelligence and adaptive control \cite{chiel1997brain,beer2000dynamical}, including the computational mechanisms underlying motor learning \cite{wolpert2011sensorimotor}. In the longer term, coupling such predictive models with artificial intelligence and robotic experimentation could also support autonomous scientific discovery by enabling iterative cycles of hypothesis generation, experimental intervention, observation, and model refinement \cite{king2009automation,sparkes2010robot,hase2019next}.

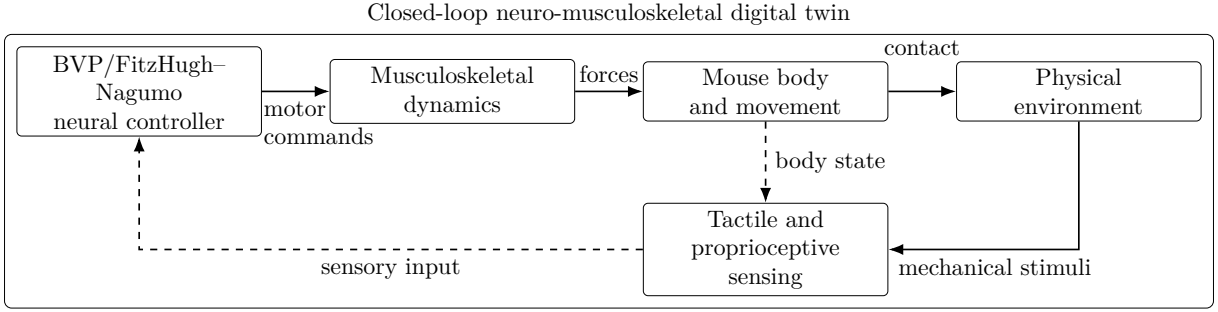
\begin{figure*}[t]
\centering
\resizebox{1.0\textwidth}{!}{%
\begin{tikzpicture}[
  node distance=10mm and 10mm,
  font=\small
]
  \node[dtwide] (neural)
    {BVP/FitzHugh--Nagumo\\neural controller};

  \node[dtwide, right=of neural] (muscle)
    {Musculoskeletal\\dynamics};

  \node[dtwide, right=of muscle] (body)
    {Mouse body\\and movement};

\node[dtwide, minimum width=30.4mm, right=of body] (env)
  {Physical\\environment};

  \node[dtwide, below=12mm of body] (sensory)
    {Tactile and\\proprioceptive sensing};

  \draw[dtarrow]
    (neural) --
    node[pos=.84, below, align=left]{motor\\commands}
    (muscle);

  \draw[dtarrow]
    (muscle) --
    node[above]{forces}
    (body);

  \draw[dtarrow]
    (body) --
    node[above=4mm]{contact}
    (env);

  \draw[dtarrow]
    (env.south) |-
    node[pos=.72, below]{mechanical stimuli}
    (sensory.east);

  \draw[dtfeedback]
    (sensory.west) -|
    node[pos=.25, below]{sensory input}
    (neural.south);

  \draw[dtfeedback]
    (body.south) --
    node[right]{body state}
    (sensory.north);

  \node[
    dtgroup,
    fit=(neural)(muscle)(body)(env)(sensory),
    label={
      [font=\small]
      above:Closed-loop neuro-musculoskeletal digital twin
    }
  ] {};

\end{tikzpicture}%
}

\caption[Conceptual architecture of the proposed mouse digital twin.]
{Conceptual architecture of the proposed neuro-musculoskeletal
mouse digital twin. Neural dynamics generate motor commands that
activate muscles and produce musculoskeletal motion
\cite{ramalingasetty2021mouse,tytell2011spikes}. Interactions between
the body and the environment generate tactile and proprioceptive
feedback that continuously modulates subsequent neural activity
\cite{proske2012proprioceptive,ijspeert2023integration}. The resulting
closed-loop system follows an embodied perspective in which adaptive
behavior emerges from interactions among the nervous system, body,
and environment
\cite{chiel1997brain,ramdya2023neuromechanics}. In the longer term,
coupling such a digital twin with robotic experimentation could support
iterative, model-guided scientific discovery
\cite{king2009automation,hase2019next}.}
\label{fig:framework}
\end{figure*}

\section*{Multimodal Anatomical Reconstruction}

A key advantage of the proposed framework is its ability to leverage multimodal imaging data for constructing anatomically realistic neuro-musculoskeletal digital twins. The feasibility of integrating complementary mouse-imaging modalities into a common three-dimensional anatomical representation has been demonstrated by atlases combining X-ray CT with registered color cryosection data \cite{dogdas2007digimouse}. Contrast-enhanced micro-CT, anatomical dissection, and digital segmentation have also been used to reconstruct mouse musculoskeletal geometry and quantify muscle architecture \cite{charles2016geometry}, providing essential parameters for biomechanical models \cite{ramalingasetty2021mouse}. In addition, genetically encoded reporters such as Scx-GFP enable the selective visualization of tendons and related connective tissues that may be difficult to distinguish using conventional structural imaging alone \cite{pryce2007transgenic}.

We therefore envision future mouse digital twins being derived from the spatial registration and integration of complementary imaging modalities, with each modality contributing distinct but mutually informative anatomical or functional information. Such multimodal integration could combine the accurate skeletal geometry provided by X-ray CT, the soft-tissue detail available from high-resolution serial-section imaging, and molecularly specific fluorescence signals from reporter animals within a unified computational representation.

X-ray computed tomography (CT), particularly high-resolution micro-CT, provides high-contrast visualization of mineralized tissues and is widely used to characterize three-dimensional bone morphology in small animals \cite{bouxein2010guidelines}. Individual bones can be segmented and labeled from volumetric CT data to reconstruct the skull, vertebral column, ribs, pelvis, and limb skeleton and to define an articulated skeletal representation \cite{khmelinskii2011articulated}. These subject-specific bone geometries provide the structural framework for specifying body segments, joints, and inertial properties in a musculoskeletal model.

Bone surfaces also provide the anatomical coordinate system on which muscle and tendon origins, insertions, and paths can be defined. However, because musculotendon attachment regions are generally not directly distinguishable in conventional CT images, they must be identified using complementary information such as contrast-enhanced micro-CT, serial-section imaging, anatomical dissection, fluorescent reporters, or registration of an anatomical atlas \cite{charles2016geometry,charles2016momentarms}. The resulting integration of skeletal geometry and musculotendon anatomy enables the construction of anatomically grounded biomechanical models.

High-resolution white-light serial-section data acquired at a
section interval of approximately 25~$\mu$m can provide detailed
color-anatomical information about soft tissues that are difficult
to distinguish using conventional CT alone. Serial block-face
cryo-imaging has been shown to generate microscopic, three-dimensional
bright-field representations of whole-mouse anatomy while maintaining
the spatial correspondence between consecutive sections
\cite{wilson2008cryo,roy2009cryo}. Moreover, registration of color
cryosection images with X-ray CT has enabled complementary skeletal
and soft-tissue information to be incorporated into a common
three-dimensional mouse atlas \cite{dogdas2007digimouse}.

Following anatomical annotation and three-dimensional segmentation,
such data can support reconstruction of muscles and other soft tissues
while preserving their spatial relationships to the skeleton.
Previous mouse limb atlases demonstrate that serial imaging combined
with tissue-specific labeling can reconstruct the three-dimensional
organization of muscles, tendons, and skeletal structures
\cite{delaurier2008limb}. When registered to the CT-derived skeleton,
these reconstructions can be used to estimate muscle paths, origins,
insertions, and other geometric constraints required for
musculoskeletal modeling \cite{charles2016geometry,charles2016momentarms}.

To incorporate tendon and other musculoskeletal connective-tissue
components, we propose using high-resolution Scx-GFP section data
acquired at a section interval comparable to that of the white-light
dataset. Scleraxis (Scx) is a basic helix--loop--helix transcription
factor expressed in tendon progenitors and in most embryonic
tendon and ligament cells, and Scx-GFP reporter mice were developed
to visualize these tissues \cite{pryce2007transgenic,murchison2007tendon}.
Scx-GFP imaging can therefore provide spatially specific information
about tendon and ligament architecture, including structures involved
in transmitting forces between muscles and bones. Scx-lineage cells
also contribute to the reproducible formation of muscle shapes and
attachment sites, further supporting the relevance of Scx-based
imaging for reconstructing musculotendinous organization
\cite{ono2023scleraxis}.

When co-registered with CT-derived skeletal geometry and white-light
soft-tissue anatomy, Scx-GFP images may facilitate more accurate
identification of tendon trajectories, myotendinous interfaces, and
muscle attachment regions that are critical for biomechanical
simulation. Multimodal anatomical atlases have previously demonstrated
the value of registering complementary CT and color serial-section
images within a shared coordinate system \cite{dogdas2007digimouse}.
The proposed extension incorporates molecularly specific fluorescence
as an additional source of anatomical information.

The integration of these complementary modalities can be formulated
as a multimodal registration and fusion problem
\cite{maintz1998registration}:

\begin{equation}
\mathcal{M}
=
\mathcal{R}
\left(
I_{\mathrm{CT}},
I_{\mathrm{WL}},
I_{\mathrm{Scx}}
\right),
\end{equation}

where $I_{\mathrm{CT}}$, $I_{\mathrm{WL}}$, and $I_{\mathrm{Scx}}$
denote the CT, white-light, and Scx-GFP image volumes, respectively,
and $\mathcal{R}$ represents a sequence of spatial registration,
segmentation, and multimodal fusion operations that produces a unified
anatomical model $\mathcal{M}$.

\begin{figure*}[t]
\centering
\resizebox{1.0\textwidth}{!}{%
\begin{tikzpicture}[
  node distance=8mm and 8mm,
  font=\small
]
  \node[dtbox] (ct)
    {X-ray CT\\bone geometry};

  \node[dtbox, below=of ct] (wl)
    {White-light sections\\soft-tissue anatomy};

  \node[dtbox, below=of wl] (scx)
    {Scx-GFP sections\\tendon-associated signal};

  \node[dtwide, right=12mm of wl] (prep)
    {Preprocessing,\\segmentation, and\\quality control};

  \node[dtwide, right=of prep] (reg)
    {Cross-modal\\registration to a\\shared coordinate frame};

  \node[dtwide, right=of reg] (fusion)
    {Unified anatomical\\representation $\mathcal{M}$};

  \node[dtbox, above right=7mm and 10mm of fusion] (skeleton)
    {Bones and joints};

  \node[dtbox, right=10mm of fusion] (muscles)
    {Muscle paths and\\attachments};

  \node[dtbox, below right=7mm and 10mm of fusion] (tendons)
    {Tendons and\\connective tissues};

  \draw[dtarrow]
    (ct.east) -- ++(4mm,0) |- (prep.west);

  \draw[dtarrow]
    (wl) -- (prep);

  \draw[dtarrow]
    (scx.east) -- ++(4mm,0) |- (prep.west);

  \draw[dtarrow]
    (prep) -- (reg);

  \draw[dtarrow]
    (reg) -- (fusion);

  \draw[dtarrow]
    (fusion.east) -- ++(4mm,0) |- (skeleton.west);

  \draw[dtarrow]
    (fusion) -- (muscles);

  \draw[dtarrow]
    (fusion.east) -- ++(4mm,0) |- (tendons.west);

\end{tikzpicture}%
}

\caption[Multimodal anatomical-reconstruction workflow.]
{Proposed multimodal reconstruction workflow. CT provides skeletal
geometry, white-light serial sections provide complementary soft-tissue
anatomy, and Scx-GFP sections provide tendon-associated fluorescence.
After preprocessing, segmentation, and cross-modal registration, the
datasets are fused into a common anatomical representation from which
simulation-ready musculoskeletal structures can be derived.}
\label{fig:multimodal-reconstruction}
\end{figure*}
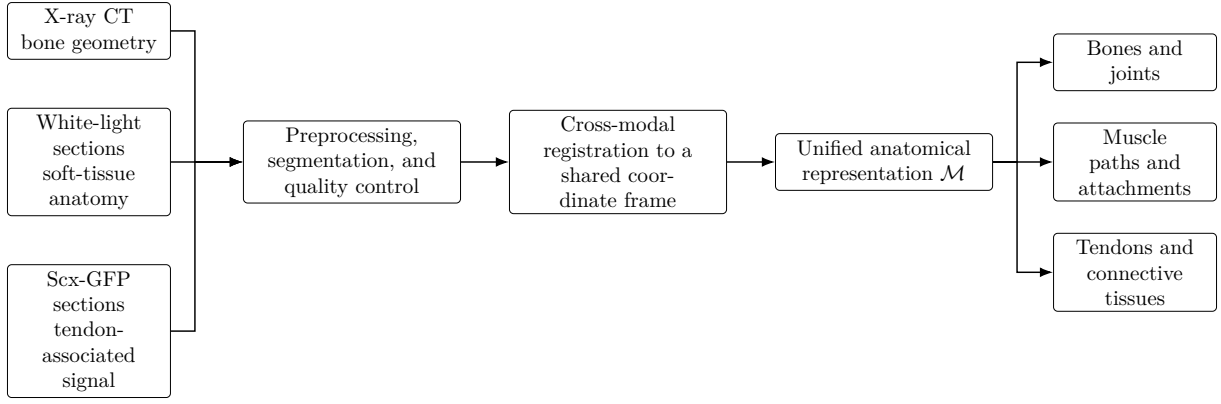

The proposed reconstruction workflow is illustrated in
Figure~\ref{fig:multimodal-reconstruction}. Each modality is first
preprocessed and segmented in its native coordinate system before
registration to a shared anatomical space. The fused representation is
then converted into simulation-ready bones, joints, muscle paths, and
tendon structures.

We argue that multimodal anatomical reconstruction will become an
essential component of future neuro-musculoskeletal digital twins.
Previous studies have demonstrated the feasibility of co-registering
X-ray CT and color cryosection data to construct three-dimensional
whole-mouse anatomical atlases \cite{dogdas2007digimouse}, while
high-resolution serial cryo-imaging can provide microscopic
bright-field representations of soft-tissue anatomy
\cite{wilson2008cryo,roy2009cryo}. Scx-GFP reporter imaging offers
complementary molecular specificity for identifying tendons,
ligaments, and related musculoskeletal interfaces
\cite{pryce2007transgenic,murchison2007tendon}.

By combining the skeletal fidelity of CT imaging, the soft-tissue
detail of white-light serial-section imaging, and the
connective-tissue information provided by Scx-GFP imaging, it may
become possible to generate detailed anatomical representations of
individual animals. Image-derived mouse models have already shown
that anatomical data can be used to define skeletal geometry,
muscle architecture, musculotendon paths, and attachment locations
for biomechanical simulation
\cite{charles2016geometry,charles2016momentarms,
ramalingasetty2021mouse}. Such anatomically grounded and
subject-specific models could provide a structural foundation for
personalized biomechanical simulation and, when coupled with neural
controllers and sensory feedback, adaptive sensorimotor simulation
\cite{ramdya2023neuromechanics,wangchen2026embodied}.

\section*{Musculoskeletal Component}

We argue that realistic body mechanics are an essential component of
future neuro-musculoskeletal digital twins. Neural activity does not
operate in isolation: the behavioral consequences of neural commands
depend on muscle properties, skeletal geometry, joint constraints,
sensory feedback, and the mechanical interaction of the body with its
environment \cite{chiel1997brain,nishikawa2007neuromechanics,
tytell2011spikes}. Conversely, body motion and muscle state generate
proprioceptive and exteroceptive feedback that continuously influences
subsequent neural activity and motor output
\cite{proske2012proprioceptive,ijspeert2023integration}.

Consequently, predictive models of animal behavior must incorporate
biomechanical representations capable of capturing muscle-driven
movement, articulated-body dynamics, contact forces, and
body--environment interactions. Existing mouse musculoskeletal models
demonstrate the feasibility of representing anatomically grounded
bones, joints, and musculotendon units for three-dimensional analysis
of movement and force generation
\cite{charles2016momentarms,ramalingasetty2021mouse}. Incorporating
such models into closed-loop neuromechanical simulations is therefore
a necessary step toward digital twins that reproduce the embodied
processes underlying animal behavior
\cite{ramdya2023neuromechanics,wangchen2026embodied}.

The proposed framework employs a whole-body mouse musculoskeletal
model implemented in simulation platforms such as OpenSim or MuJoCo.
OpenSim supports anatomically grounded modeling and simulation of
musculoskeletal dynamics and neuromuscular control
\cite{delp2007opensim,seth2018opensim}, whereas MuJoCo provides
efficient generalized-coordinate simulation of articulated systems,
actuators, and environmental contacts
\cite{todorov2012mujoco}. A whole-body mouse model comprising skeletal
segments, articulated joints, and Hill-type musculotendon units has
already demonstrated the feasibility of this approach
\cite{ramalingasetty2021mouse}.

Rather than serving merely as a passive mechanical substrate, the
musculoskeletal system actively transforms neural excitation into
muscle force and movement. This transformation depends on nonlinear
muscle force--length and force--velocity properties, tendon compliance,
skeletal geometry, and external loading
\cite{zajac1989muscle,tytell2011spikes}. Movement and contact, in turn,
alter proprioceptive and exteroceptive inputs that influence subsequent
neural activity, forming a closed sensorimotor loop
\cite{proske2012proprioceptive,ijspeert2023integration}.

At a conceptual level, the biomechanical state can be represented as

\begin{equation}
x_m =
\left(
q,\dot{q},a,l_f,\dot{l}_f
\right),
\end{equation}

where $q$ and $\dot{q}$ denote generalized positions and velocities,
$a$ denotes muscle activation, and $l_f$ and $\dot{l}_f$ denote muscle
fiber lengths and shortening or lengthening velocities, respectively.
Depending on the adopted musculotendon formulation, tendon lengths or
other internal muscle states may also be included
\cite{zajac1989muscle}.

The motion of the articulated body can be written generally as

\begin{equation}
M(q)\ddot{q}
+
C(q,\dot{q})
+
G(q)
=
\tau_{\mathrm{muscle}}
+
J_c(q)^{\mathsf T} f_c
+
\tau_{\mathrm{ext}},
\end{equation}

where $M(q)$ is the mass matrix, $C(q,\dot{q})$ represents
velocity-dependent inertial terms, $G(q)$ represents gravitational
forces, $\tau_{\mathrm{muscle}}$ denotes generalized forces generated
by musculotendon actuators, $f_c$ denotes environmental contact forces,
and $J_c(q)^{\mathsf T}$ maps those forces into generalized coordinates.
The term $\tau_{\mathrm{ext}}$ represents any additional externally
applied generalized forces. Movement therefore emerges from the coupled
interaction of muscle dynamics, articulated-body mechanics, and
environmental contact \cite{seth2018opensim,todorov2012mujoco}.

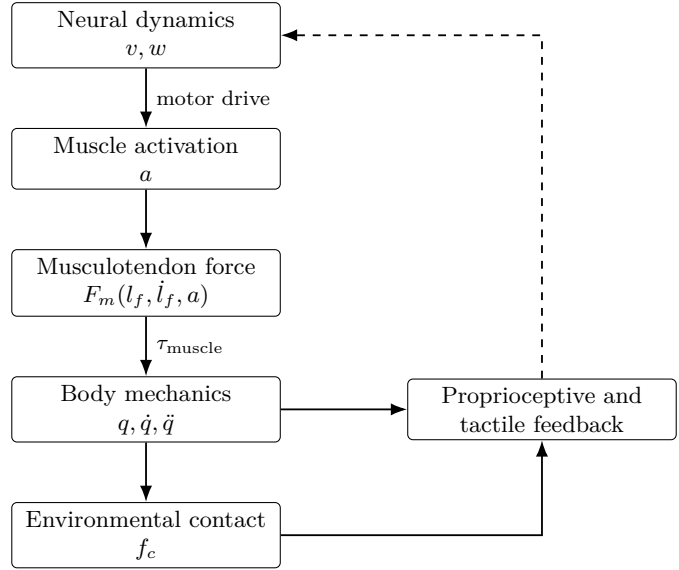
\begin{figure}[t]
\centering
\resizebox{0.55\columnwidth}{!}{%
\begin{tikzpicture}[node distance=8mm, font=\footnotesize]
  \node[dtwide] (controller) {Neural dynamics\\$v,w$};
  \node[dtwide, below=of controller] (activation) {Muscle activation\\$a$};
  \node[dtwide, below=of activation] (muscleforce)
    {Musculotendon force\\$F_m(l_f,\dot l_f,a)$};
  \node[dtwide, below=of muscleforce] (mechanics)
    {Body mechanics\\$q,\dot q,\ddot q$};
  \node[dtwide, below=of mechanics] (contact)
    {Environmental contact\\$f_c$};
  \node[dtwide, right=17mm of mechanics] (feedback)
    {Proprioceptive and\\tactile feedback};

  \draw[dtarrow] (controller)
    -- node[right, font=\scriptsize]{motor drive} (activation);
  \draw[dtarrow] (activation) -- (muscleforce);
  \draw[dtarrow] (muscleforce)
    -- node[right, font=\scriptsize]{$\tau_{\mathrm{muscle}}$} (mechanics);
  \draw[dtarrow] (mechanics) -- (contact);
  \draw[dtarrow] (contact.east) -| (feedback.south);
  \draw[dtarrow] (mechanics.east) -- (feedback.west);
  \draw[dtfeedback] (feedback.north) |- (controller.east);
\end{tikzpicture}}
\caption[Coupling between neural dynamics and body mechanics.]
{Information and force exchange in the neuromechanical simulation.
Neural states drive muscle activation and musculotendon force, which
generate articulated-body motion and environmental contact. Body and
contact states are transformed into proprioceptive and tactile feedback
that modulates subsequent neural dynamics.}
\label{fig:neuromechanical-coupling}
\end{figure}

Figure~\ref{fig:neuromechanical-coupling} summarizes the variables
exchanged across the neural, muscular, mechanical, and sensory
subsystems. It emphasizes that the musculoskeletal model is both an
actuated physical plant and a source of sensory state variables.

Importantly, the purpose of biomechanical simulation is not simply
to reproduce observed movement. Rather, we view the body as a
computational medium through which neural control, sensory feedback,
and environmental interactions become tightly coupled. This view is
consistent with embodied and dynamical-systems accounts in which
behavior emerges from the recurrent interaction of the nervous system,
body, and environment \cite{chiel1997brain,beer2000dynamical,
tytell2011spikes}. It is also closely related to the concept of
morphological computation, according to which the geometry, compliance,
inertia, and intrinsic dynamics of the body contribute directly to
behavioral control and information processing
\cite{pfeifer2009morphological,zahedi2013quantifying}.

This perspective suggests that realistic musculoskeletal models will
provide a critical foundation for predictive animal digital twins.
By coupling neural controllers to anatomically and mechanically
grounded bodies within simulated environments, neuromechanical digital
twins may reveal otherwise inaccessible biophysical variables, permit
controlled perturbations, and generate experimentally testable
hypotheses \cite{ramdya2023neuromechanics,wangchen2026embodied}.
Such models could support prediction and comparison of interventions;
however, their use for intervention planning would require
subject-specific calibration and prospective experimental validation.
In the longer term, integration with artificial intelligence and
robotic experimentation could enable iterative cycles of hypothesis
generation, experimental selection, observation, and model refinement,
thereby contributing to autonomous scientific discovery
\cite{king2009automation,sparkes2010robot,hase2019next}.

\section*{Neural Dynamics Using the Bonhoeffer--van der Pol Model}

Rather than beginning with large networks of biophysically detailed
spiking neurons, we propose the Bonhoeffer--van der Pol (BVP) model as
a reduced-order representation of excitable neural and sensorimotor
dynamics. FitzHugh introduced the BVP model as a two-variable
simplification of the dynamical behavior exhibited by the
Hodgkin--Huxley nerve-membrane equations, and Nagumo and colleagues
subsequently implemented a closely related system as an electrical
circuit for nerve-pulse propagation
\cite{fitzhugh1961impulses,nagumo1962active}. The model captures
qualitative phenomena including excitability, threshold behavior,
recovery, refractoriness, and self-sustained oscillation using only
two state variables. Its low dimensionality makes it attractive for
large coupled simulations in which neural dynamics must be integrated
with biomechanical and sensory processes.

For each neuron or reduced neural module, we use the nondimensional
form

\begin{equation}
\frac{dv_i}{dt}
=
v_i-\frac{v_i^3}{3}-w_i+I_i
+\sum_j K_{ij}H(v_j,v_i),
\label{eq:fhn1}
\end{equation}

\begin{equation}
\frac{dw_i}{dt}
=
\epsilon_i\left(v_i+a_i-b_iw_i\right),
\label{eq:fhn2}
\end{equation}

where \(v_i\) is a fast excitation variable commonly interpreted as
a reduced membrane potential, \(w_i\) is a slow recovery variable,
and \(I_i\) represents external input, including sensory or descending
drive. The parameters \(a_i\), \(b_i\), and \(\epsilon_i\) determine
the nullclines, excitability, and relative time scales of the fast and
slow variables. The optional term \(K_{ij}H(v_j,v_i)\) represents
coupling from module \(j\) to module \(i\); its precise form must be
selected according to the intended neural interaction.

The BVP model occupies a useful position in the hierarchy of neural
representations. Detailed conductance-based models emphasize cellular
and ionic mechanisms, whereas many engineering controllers primarily
optimize task performance. The BVP framework instead preserves a
compact phase-space description of excitability and oscillation
\cite{fitzhugh1961impulses}. This reduction offers computational
advantages, although it does not reproduce the cellular diversity,
synaptic physiology, or detailed spike timing of larger biophysical
networks. Other reduced neuron models likewise demonstrate that
low-dimensional dynamics can provide a useful compromise between
biological expressiveness and computational cost
\cite{izhikevich2003simple}.

At the network level, coupled nonlinear oscillators provide established
models of central pattern generators and can generate coordinated
rhythmic motor commands for animal and robotic locomotion
\cite{ijspeert2008central,righetti2006programmable}. Sensory input can
entrain or modulate these oscillators, enabling coordination to emerge
through closed-loop interactions among the controller, body, and
environment \cite{ijspeert2023integration}. Thus, locomotor rhythm and
motor coordination can plausibly be represented as collective dynamics
of coupled BVP modules. Tactile sensing and adaptive behavior, however,
should not be identified with oscillation alone: tactile receptors,
sensory encoding, network connectivity, coupling laws, and adaptation
mechanisms must be specified explicitly.

This dynamical-systems viewpoint aligns naturally with embodied
intelligence, in which behavior arises from continuous interactions
among neural activity, body mechanics, sensory feedback, and the
environment \cite{chiel1997brain,beer2000dynamical}. A shared
differential-equation framework can couple BVP neural modules to
musculoskeletal states and sensory inputs without treating these
subsystems as independent. Moreover, adaptive-oscillator studies have
shown that intrinsic frequencies and coupling parameters can be learned
from periodic inputs, demonstrating one possible route to online
adaptation \cite{righetti2006hebbian}. We therefore propose BVP
dynamics as a computationally economical candidate for scalable
neuro-musculoskeletal digital twins supporting closed-loop control and
adaptive learning. Their extension to autonomous experimentation would
additionally require model updating, uncertainty estimation,
experimental-design algorithms, and interfaces to robotic systems
\cite{king2009automation,hase2019next}.

\section*{Tactile Sensation Model}

We argue that tactile sensation is a critical but frequently simplified
component of animal digital twins. Considerable progress has been made
in modeling neural circuits and musculoskeletal mechanics, but
neuromechanical simulations often represent sensory inputs using
prescribed signals or simplified feedback laws. In biological motor
systems, by contrast, cutaneous and proprioceptive mechanosensory
pathways are integrated with descending commands and local motor
circuits to support precise and adaptive movement
\cite{abraira2013touch,goulding2025sensory}. Behavior therefore emerges
through continuous interactions among the nervous system, body, and
environment rather than from feedforward motor commands alone
\cite{chiel1997brain,tytell2011spikes}.

In the proposed framework, tactile sensing serves as a major interface
between the simulated physical environment and the neural controller.
Contacts at the forepaws and hindpaws activate specialized
mechanoreceptor populations in glabrous skin; the mouse forepaw, in
particular, has dense and physiologically specialized mechanoreceptor
innervation suited to tactile exploration
\cite{walcher2018mechanoreceptors}. Contacts involving hairy skin and
the body surface activate other classes of low-threshold
mechanoreceptors with distinct spatial and temporal response
properties \cite{zimmerman2014gentle}. Whisker contact constitutes a
specialized active-sensing system in which forces and bending moments
at the whisker base are converted into neural signals by receptors in
the follicle \cite{diamond2008whisker,bush2016whisker}. These tactile
signals can modify ongoing neural activity and subsequent motor action.
Tactile feedback should therefore be treated as an integral component
of the closed sensorimotor loop rather than as an auxiliary output.

At a conceptual level, sensory transduction can initially be
approximated as a mapping from local mechanical interaction to neural
input. For a receptor or sensory unit \(i\), a simple static model is

\begin{equation}
I_{\mathrm{sens},i}
=
k_i\,[F_{n,i}-F_{\mathrm{th},i}]_{+},
\label{eq:tactile_static}
\end{equation}

where \(F_{n,i}\) is the local normal contact force,
\(F_{\mathrm{th},i}\) is a response threshold, \(k_i\) is a gain, and
\([z]_{+}=\max(0,z)\). This expression is useful as a minimal
implementation, but it does not reproduce the adaptation, directional
selectivity, receptive fields, or temporal filtering of biological
mechanoreceptors \cite{abraira2013touch,walcher2018mechanoreceptors}.

A more general dynamical formulation is

\begin{equation}
\tau_i\frac{dI_{\mathrm{sens},i}}{dt}
=
-I_{\mathrm{sens},i}
+
\phi_i
\left(
F_{n,i},
F_{t,i},
\dot{F}_{n,i},
\dot{F}_{t,i},
x_i,
s_i
\right),
\label{eq:tactile_dynamic}
\end{equation}

where \(F_{n,i}\) and \(F_{t,i}\) denote local normal and tangential
contact forces, their derivatives describe dynamic loading,
\(x_i\) denotes contact position or receptive-field location, and
\(s_i\) represents additional receptor-specific state variables.
The function \(\phi_i\) can be selected to represent slowly adapting,
rapidly adapting, direction-selective, or whisker-associated sensory
units. For whiskers, base bending moment and angular deflection may be
more appropriate inputs than surface normal and tangential forces
\cite{bush2016whisker}.

\begin{figure}[t]
\centering
\resizebox{0.5\columnwidth}{!}{%
\begin{tikzpicture}[node distance=8mm, font=\small]
  \node[dtwide] (contact) {Body--environment contact\\$F_n,F_t,\dot F,x$};
  \node[dtwide, below=of contact] (local) {Local tissue and whisker\\mechanics};
  \node[dtwide, below=of local] (receptor) {Receptor-specific filtering\\threshold, gain, adaptation};
  \node[dtwide, below=of receptor] (sensory) {Sensory drive\\$I_{\mathrm{sens},i}$};
  \node[dtwide, below=of sensory] (neural) {BVP/FitzHugh--Nagumo\\neural modules};
  \draw[dtarrow] (contact) -- (local);
  \draw[dtarrow] (local) -- (receptor);
  \draw[dtarrow] (receptor) -- (sensory);
  \draw[dtarrow] (sensory) -- (neural);
  \node[dtbox, right=16mm of receptor] (classes)
    {Slowly adapting\\Rapidly adapting\\Direction selective\\Whisker associated};
  \draw[dtfeedback] (classes.west) -- (receptor.east);
\end{tikzpicture}}
\caption[Mechanical-to-neural tactile transduction.]
{Conceptual tactile-sensation pipeline. Contact forces, loading rates,
positions, and whisker mechanics are processed by receptor-specific
thresholding and temporal filtering. The resulting sensory drive enters
the neural dynamics and thereby closes the sensorimotor loop.}
\label{fig:tactile-transduction}
\end{figure}
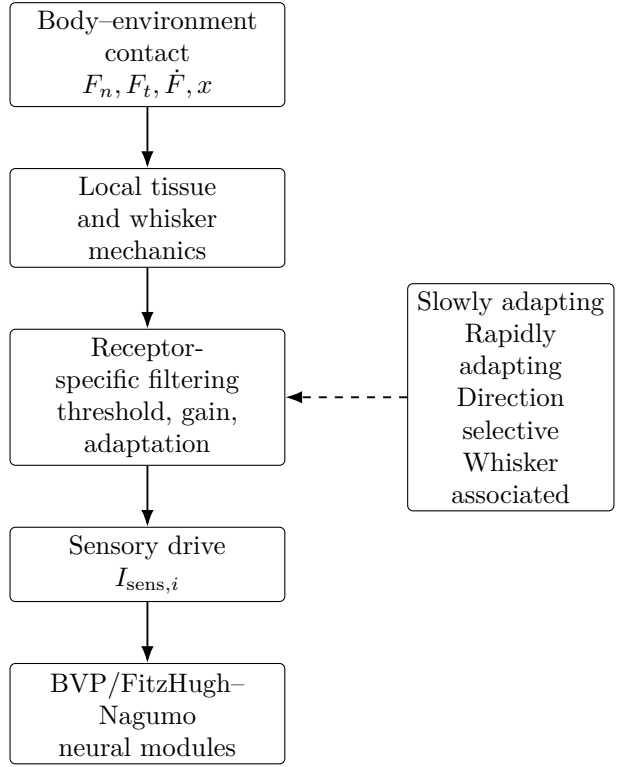

Figure~\ref{fig:tactile-transduction} shows how local contact mechanics
are converted into receptor-specific neural input. Keeping these stages
explicit permits alternative receptor models to be evaluated without
changing the mechanical simulation or neural controller.

The essential modeling principle is that simulated environmental
interactions directly and causally modify neural dynamics.
Mechanosensory feedback contributes to corrective motor responses,
locomotor adaptation, and context-dependent control
\cite{ijspeert2023integration,goulding2025sensory}. Whisker-mediated
touch additionally illustrates active sensing, in which an animal
moves its sensory apparatus to acquire task-relevant information
\cite{prescott2011active}. Incorporating these feedback pathways
therefore transforms the digital twin from a feedforward movement
generator into an embodied agent that can perceive and respond to its
simulated surroundings.

We propose that future neuro-musculoskeletal digital twins should treat
tactile sensation as a first-class component of the modeling
framework. Tight coupling among contact mechanics, sensory
transduction, neural dynamics, and movement could provide a foundation
for studying embodied intelligence, active sensing, and adaptive
behavior. Extension to autonomous scientific experimentation would
additionally require experimental-design algorithms, model updating,
uncertainty quantification, and interfaces to physical measurement and
robotic systems \cite{king2009automation,hase2019next}.

\section*{Closed-Loop Sensorimotor Integration}

A central premise of this position paper is that animal behavior cannot
be understood solely from neural activity, musculoskeletal mechanics,
or sensory processing considered in isolation. Rather, behavior emerges
from continuous interactions among the nervous system, body, and
environment within a closed-loop sensorimotor system
\cite{chiel1997brain,nishikawa2007neuromechanics,
tytell2011spikes}. This perspective is consistent with dynamical
accounts of cognition and sensorimotor behavior, which emphasize
reciprocal causation between an animal's actions and its sensory inputs
\cite{beer2000dynamical,buhrmann2013dynamical}.

In the proposed framework, neural dynamics generate motor commands that
activate muscles and produce movement through the musculoskeletal
system. Muscle and tendon properties, skeletal geometry, and external
loading transform these commands into bodily motion
\cite{zajac1989muscle,ramalingasetty2021mouse}. The resulting motion
changes the animal's interaction with its surroundings, generating
tactile and proprioceptive signals that feed back into the neural
controller. These signals contribute to the timing, coordination, and
adaptation of subsequent motor output
\cite{proske2012proprioceptive,ijspeert2023integration,
goulding2025sensory}. Consequently, perception and action are not
independent stages but mutually dependent components of an ongoing
dynamical cycle.

This closed-loop relationship can be summarized as

\begin{equation}
\begin{aligned}
\text{Sensation}
&\rightarrow
\text{Neural Dynamics}
\rightarrow
\text{Musculoskeletal Action} \\
&\rightarrow
\text{Environmental Interaction}
\rightarrow
\text{Sensation}.
\end{aligned}
\label{eq:closed_loop}
\end{equation}

Although this diagram is shown as a sequence, the underlying system is
continuous and may contain multiple concurrent feedback pathways
operating at different spatial and temporal scales. These include rapid
mechanical responses, spinal sensorimotor loops, rhythmic neural
circuits, and higher-level adaptive control
\cite{tytell2011spikes,ijspeert2008central,
goulding2025sensory}.

We argue that such sensorimotor coupling is a defining organizational
principle of adaptive biological behavior and should therefore be a
fundamental design principle of future neuro-musculoskeletal digital
twins. Unlike simulations based exclusively on prescribed trajectories
or predetermined sensory inputs, closed-loop models allow movements and
sensory states to arise jointly through interactions among the neural
controller, biomechanical body, and environment
\cite{chiel1997brain,ramdya2023neuromechanics}. This capability is
important for studying adaptive locomotion
\cite{ijspeert2023integration}, active sensing
\cite{prescott2011active}, motor learning
\cite{wolpert2011sensorimotor}, and embodied behavioral control
\cite{beer2000dynamical,buhrmann2013dynamical}.

Furthermore, closed-loop integration provides a natural foundation for
coupling digital twins with artificial intelligence and robotic
experimentation platforms. A digital twin could assimilate behavioral
and sensory observations, update its internal state or parameters,
predict the outcomes of candidate interventions, and use subsequent
measurements to refine those predictions. Neuromechanical models can
thereby function not only as movement simulators but also as tools for
generating experimentally testable hypotheses
\cite{wangchen2026embodied}. In the longer term, combining such models
with artificial intelligence, experimental-design algorithms, and
robotic systems could support adaptive control, model-guided hypothesis
generation, and real-time prediction of behavioral responses to
experimental interventions
\cite{king2009automation,sparkes2010robot,hase2019next}.

\section*{Potential Applications}

The proposed neuro-musculoskeletal digital twin extends beyond a
conventional biomechanical simulation framework. By integrating neural
dynamics, musculoskeletal mechanics, tactile sensing, and environmental
interactions within a closed-loop architecture, the model may serve as a
platform for investigating embodied intelligence across multiple domains
\cite{ramdya2023neuromechanics,ijspeert2023integration}.
We envision applications ranging from fundamental neuroscience and motor
control research to translational modeling, brain--machine interfaces, and
autonomous scientific discovery. Representative application areas are
summarized in Table~\ref{tab:applications}.

\begin{table}[t]
\centering
\caption{Potential applications of the proposed neuro-musculoskeletal
mouse digital twin.}
\label{tab:applications}

\small
\renewcommand{\arraystretch}{1.15}

\begin{tabularx}{\linewidth}{
  @{}
  >{\raggedright\arraybackslash}p{0.24\linewidth}
  >{\raggedright\arraybackslash}X
  @{}
}
\toprule
\textbf{Application} & \textbf{Purpose} \\
\midrule

Locomotion &
Study the emergence of locomotion and postural control from interactions
among neural control, musculoskeletal mechanics, and environmental
contacts
\cite{ramdya2023neuromechanics,ijspeert2023integration}. \\

\addlinespace
Active sensing &
Investigate whisker-mediated tactile exploration and the coupling among
sensing, movement, and environmental perception
\cite{diamond2008whisker,prescott2011active}. \\

\addlinespace
Neurological conditions &
Represent disease- or injury-associated alterations in neural control and
biomechanics, such as those occurring in parkinsonian movement disorders
or spinal cord injury
\cite{bonanno2025neural}. \\

\addlinespace
Brain--machine interfaces &
Explore neural-decoding, sensory-feedback, and stimulation strategies
under controlled computational assumptions
\cite{chaudhary2016brain}. \\

\addlinespace
Autonomous robotic science &
Support model-guided hypothesis generation, experiment selection, and
iterative model refinement
\cite{king2009automation,sparkes2010robot,hase2019next}. \\

\bottomrule
\end{tabularx}
\end{table}

Among these applications, locomotion and active sensing provide
particularly relevant opportunities to investigate how behavior emerges
from interactions among neural control, body mechanics, and sensory
feedback
\cite{ramdya2023neuromechanics,diamond2008whisker,prescott2011active}.
Because locomotor and exploratory behaviors depend on recurrent exchanges
between neural commands, mechanical dynamics, and sensory consequences,
they provide natural test cases for evaluating the predictive capacity of
closed-loop digital twins.

In translational contexts, disease- or injury-specific modifications of
neural, sensory, or musculoskeletal parameters could be used to investigate
candidate mechanisms and prioritize intervention hypotheses
\cite{bonanno2025neural,saxby2023digital}. Such simulations should be
regarded as complements to, rather than replacements for, animal
experiments: their predictive utility will depend on calibration against
experimental data and prospective biological validation. Integration with
brain--machine interfaces could similarly provide controlled test beds for
exploring neural decoding, artificial sensory feedback, and stimulation
strategies before their evaluation in vivo
\cite{chaudhary2016brain}.

Looking further ahead, we propose that neuro-musculoskeletal digital twins
may become components of autonomous scientific platforms. Robotic-science
systems have already demonstrated that machine-readable models can be
combined with automated experimentation to generate hypotheses, select
experiments, interpret observations, and iteratively refine scientific
models
\cite{king2009automation,sparkes2010robot,hase2019next}. Coupling these
capabilities to a neuro-musculoskeletal digital twin could allow the model
to assimilate behavioral and physiological observations, update its
parameters, identify informative experiments, and predict responses to
candidate interventions. This would represent a transition from passive
simulation toward model-guided scientific discovery, although achieving
fully autonomous operation will require robust uncertainty quantification,
experimental validation, and human oversight.

\section*{Why Bonhoeffer--van der Pol?}

The Bonhoeffer--van der Pol (BVP) model provides a common
dynamical language for describing interactions among neural activity,
body mechanics, sensation, and environmental feedback. FitzHugh
introduced the BVP equations as a reduced two-variable representation
of excitability and refractoriness in the Hodgkin--Huxley system
\cite{fitzhugh1961impulses,hodgkin1952quantitative}. Following the
electronic implementation developed by Nagumo and colleagues, closely
related formulations became widely known as the FitzHugh--Nagumo
model \cite{nagumo1962active}. Hereafter, we use the term
BVP/FitzHugh--Nagumo model to emphasize this historical and
mathematical relationship.

\begin{table}[t]
\centering
\caption[Comparison of candidate neural-control models.]
{Qualitative comparison of candidate neural-control models for
neuro-musculoskeletal digital twins. Computational cost and suitability
depend on the implementation, network size, and biological phenomena
being modeled \cite{izhikevich2004which}.}
\label{tab:fhn}

\small
\setlength{\tabcolsep}{3.5pt}
\renewcommand{\arraystretch}{1.15}

\begin{tabularx}{\linewidth}{
  @{}
  >{\raggedright\arraybackslash}p{0.26\linewidth}
  >{\centering\arraybackslash}p{0.15\linewidth}
  >{\centering\arraybackslash}p{0.16\linewidth}
  >{\raggedright\arraybackslash}X
  @{}
}
\toprule
\textbf{Model} &
\textbf{Biological realism} &
\textbf{Computational cost} &
\textbf{Suitability} \\
\midrule

Hodgkin--Huxley
\cite{hodgkin1952quantitative} &
High &
Very high &
Limited for large-scale, real-time systems. \\

\addlinespace
Spiking neural networks
\cite{izhikevich2004which} &
Moderate--high &
Moderate--high &
Moderate. \\

\addlinespace
BVP/FitzHugh--Nagumo
\cite{fitzhugh1961impulses,nagumo1962active} &
Moderate &
Low &
High for reduced-order neural dynamics. \\

\addlinespace
PID controller
\cite{astrom2001future} &
Low &
Very low &
High for engineering control, but limited in its ability to represent
neural dynamics. \\

\bottomrule
\end{tabularx}
\end{table}

For simulations involving hundreds of muscles, sensory receptors, and
neural-control modules, the BVP/FitzHugh--Nagumo model offers a
practical balance between dynamical richness and computational
scalability. Unlike the biophysically detailed Hodgkin--Huxley model,
it represents excitation and recovery using only two state variables
while retaining threshold behavior, refractoriness, oscillations, and
excitability \cite{fitzhugh1961impulses,hodgkin1952quantitative}.
These properties make it suitable for studying how distributed neural
modules interact with muscles, sensory feedback, and environmental
contacts in closed-loop simulations.

This advantage should not, however, be interpreted as universal
superiority. Detailed conductance-based models remain preferable when
ion-channel mechanisms or cellular electrophysiology are central to the
research question, whereas spiking-network models are more appropriate
when spike timing, synaptic plasticity, or population coding must be
represented explicitly \cite{izhikevich2004which}. The principal value
of the BVP/FitzHugh--Nagumo framework is therefore its suitability as a
reduced-order dynamical model for large, interactive, and potentially
real-time neuro-musculoskeletal simulations.

\section*{Future Directions}

The framework proposed in this position paper represents an initial step
toward predictive neuro-musculoskeletal digital twins capable of
capturing interactions among neural dynamics, biomechanics, sensation,
and environmental feedback. While the present formulation emphasizes
the computationally tractable integration of FitzHugh--Nagumo neural
dynamics, musculoskeletal simulation, and tactile sensing, substantial
opportunities remain for extending its biological realism, predictive
capacity, and scientific utility
\cite{ramalingasetty2021mouse,saxby2023digital}.

We envision future developments that integrate advances in
neuroscience, biomechanics, artificial intelligence, and robotics.
In particular, subject-specific anatomical reconstruction, large-scale
neural recording, adaptive learning, data-driven parameter estimation,
and autonomous experimentation may enable increasingly realistic and
continuously updated digital representations of behaving animals.
Representative research directions are summarized in
Table~\ref{tab:future}.

\begin{table}[t]
\centering
\caption{Future directions for neuro-musculoskeletal digital twins.}
\label{tab:future}
\small
\renewcommand{\arraystretch}{1.15}
\begin{tabularx}{\linewidth}{
  @{}
  >{\raggedright\arraybackslash}p{0.24\linewidth}
  >{\raggedright\arraybackslash}X
  @{}
}
\toprule
\textbf{Direction} & \textbf{Potential impact} \\
\midrule

Subject-specific anatomy &
Integration of micro-CT, histological, and molecular imaging data to
construct anatomically grounded, individual-specific musculoskeletal
models \cite{ramalingasetty2021mouse,saxby2023digital}. \\

\addlinespace
Whole-brain recording and modeling &
Integration of large-scale neural recordings with brain-network models
and biomechanical simulations
\cite{steinmetz2021neuropixels,sanzleon2013virtual}. \\

\addlinespace
Reinforcement learning &
Adaptive optimization of sensorimotor policies and investigation of
motor-learning processes in muscle-driven simulations
\cite{song2021deep}. \\

\addlinespace
Neuromorphic hardware &
Low-latency, energy-efficient implementation of neural dynamics for
real-time closed-loop simulation and control
\cite{davies2018loihi}. \\

\addlinespace
Data-driven parameter estimation &
Automated calibration of neural, sensory, and musculoskeletal parameters,
together with assessment of parameter identifiability and predictive
uncertainty \cite{villaverde2019benchmarking}. \\

\addlinespace
Robotic validation platforms &
Iterative refinement of digital twins through automated behavioral
experiments, model-based hypothesis generation, and closed-loop testing
\cite{king2009automation,sparkes2010robot}. \\

\bottomrule
\end{tabularx}
\end{table}

Subject-specific reconstruction will require methods for registering and
fusing skeletal, muscular, connective-tissue, and neural datasets while
quantifying uncertainty introduced by imaging, segmentation, and
cross-modal registration. Existing whole-body mouse musculoskeletal
models provide an important starting point, but individualized twins
will require direct estimation of geometry, inertial properties, muscle
paths, and musculotendon parameters from each animal
\cite{ramalingasetty2021mouse}.

At the neural level, high-density recording technologies such as
Neuropixels make it possible to record large neuronal populations over
extended periods in freely behaving small animals
\cite{steinmetz2021neuropixels}. These data could constrain regional or
network-level models of neural dynamics, including multiscale
whole-brain modeling frameworks
\cite{sanzleon2013virtual}. However, large-scale recordings do not
constitute a complete measurement of whole-brain activity; unobserved
states will still need to be inferred through explicit observation
models and data-assimilation methods.

Reinforcement learning may provide a complementary approach for
discovering sensorimotor control policies in high-dimensional
musculoskeletal systems. Previous studies have demonstrated that deep
reinforcement learning can generate locomotor control in
neuromechanical simulations, although physiological plausibility
depends strongly on the reward function, model structure, training
data, and biomechanical constraints \cite{song2021deep}. Future work
could combine learned policies with interpretable dynamical controllers
such as coupled FitzHugh--Nagumo modules, allowing adaptive behavior
without completely abandoning mechanistic interpretability.

Neuromorphic processors may further support low-latency implementation
of large neural controllers by exploiting event-driven computation and
local learning mechanisms \cite{davies2018loihi}. Nevertheless, the
computational advantages of a particular neuromorphic platform must be
demonstrated for the selected neural equations; hardware designed
primarily for spiking neural networks will not necessarily accelerate
continuous FitzHugh--Nagumo dynamics without an appropriate mapping or
discretization strategy.

A further priority is automated parameter estimation from experimental
observations. Neural, biomechanical, and sensory models may contain
many parameters that cannot be uniquely determined from a single
experiment. Consequently, model calibration should be accompanied by
sensitivity analysis, identifiability assessment, cross-validation, and
uncertainty quantification
\cite{villaverde2019benchmarking}. This is essential if predictions
from the digital twin are to be used for selecting interventions or
designing subsequent experiments.

Finally, robotic experimentation could close the loop between
simulation and empirical validation. Robot-scientist systems have
demonstrated that automated instruments can execute experiments,
interpret results, and select subsequent hypotheses within restricted
experimental domains
\cite{king2009automation,sparkes2010robot}. Extending this approach to
behavioral neuroscience could allow a digital twin to identify
informative perturbations, compare predicted and observed responses,
and update its internal parameters. Such systems should initially
operate as human-supervised scientific assistants, with explicit
constraints for animal welfare, experimental safety, uncertainty, and
interpretability.

\subsection*{Reinforcement Learning and Adaptive Behavior}

Among the future directions outlined above, reinforcement learning (RL)
may provide a particularly powerful mechanism for enabling adaptive
behavior in neuro-musculoskeletal digital twins. RL formalizes how an
agent can improve its behavior through trial-and-error interactions with
a dynamic environment and delayed evaluative feedback
\cite{kaelbling1996reinforcement,kober2013reinforcement}. This framework
has conceptual connections to biological reward-based learning, although
an engineered RL algorithm should not be interpreted as a complete model
of the biological mechanisms underlying learning
\cite{schultz1997neural}.

In the proposed framework, the digital twin can be treated as an embodied
agent whose state is determined by neural activity, musculoskeletal
configuration, sensory feedback, and relevant properties of the
environment. At time $t$, the agent receives a state or observation
$s_t$, selects an action $a_t$ according to a policy
$\pi(a_t \mid s_t)$, and subsequently receives a reward $r_t$ and a new
observation. Motor actions generated by the neural controller influence
the environment through body movement, whereas tactile and
proprioceptive feedback provide information about the consequences of
those actions.

Because an animal or digital twin generally has incomplete access to the
full state of its body and environment, the problem may be more
appropriately formulated as a partially observable Markov decision
process. In this case, the policy may depend on an internal state or a
history of sensory observations rather than only on the instantaneous
observation $s_t$ \cite{kaelbling1996reinforcement}.

A reward function may be defined to encourage behaviors such as stable
locomotion, obstacle avoidance, target reaching, robustness to
perturbations, or reduced energetic expenditure. At a conceptual level,
the learning objective can be expressed as

\begin{equation}
\pi^{*}
=
\arg\max_{\pi}
\mathbb{E}_{\tau \sim \pi}
\left[
\sum_{t=0}^{T}
\gamma^{t} r_t
\right],
\label{eq:rl_objective}
\end{equation}

where $\pi$ denotes the control policy, $\tau$ is a trajectory generated
by interactions between the policy and the environment, $r_t$ is the
reward at time step $t$, and $\gamma \in [0,1]$ is a discount factor.
Through repeated simulated interactions, the agent updates its policy to
increase expected cumulative reward.

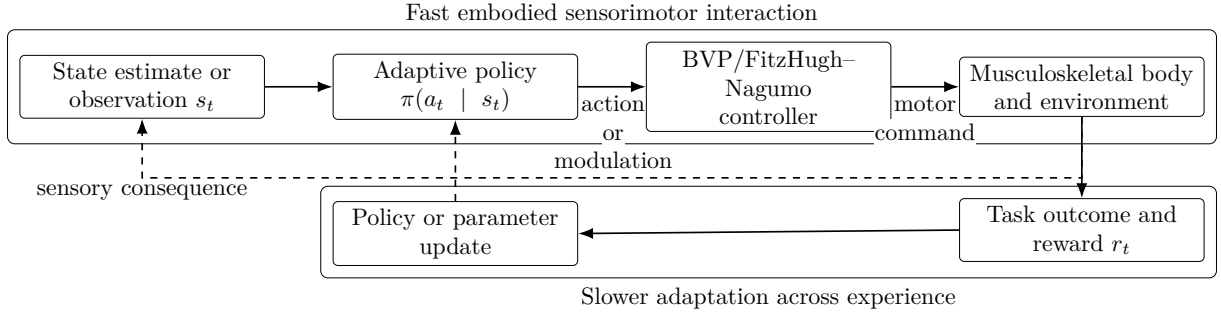
\begin{figure*}[t]
\centering

\begin{adjustbox}{max width=1.0\textwidth}
\begin{tikzpicture}[
  node distance=10mm and 10mm,
  font=\small
]
  \node[dtwide] (obs)
    {State estimate or\\observation $s_t$};

  \node[dtwide, right=of obs] (policy)
    {Adaptive policy\\$\pi(a_t\mid s_t)$};

  \node[dtwide, right=of policy] (controller)
    {BVP/FitzHugh--Nagumo\\controller};

  \node[dtwide, right=of controller] (body)
    {Musculoskeletal body\\and environment};

  \node[dtwide, below=12mm of body] (reward)
    {Task outcome and\\reward $r_t$};

  \node[dtwide, below=12mm of policy] (update)
    {Policy or parameter\\update};

  \node[
    dtgroup,
    fit=(obs)(policy)(controller)(body),
    label={
      [font=\small]
      above:Fast embodied sensorimotor interaction
    }
  ] {};

\draw[dtarrow]
  (policy) --
  node[pos=.52, below, align=center]
  {action\\
   {\setlength{\fboxsep}{0.25pt}\colorbox{white}{\strut or}}\\
   modulation}
  (controller);

\draw[dtarrow]
  (controller) --
  node[below, align=center]
  {motor\\
   {\setlength{\fboxsep}{0.25pt}\colorbox{white}{\strut command}}}
  (body);

  \node[
    dtgroup,
    fit=(reward)(update),
    label={
      [font=\small]
      below:Slower adaptation across experience
    }
  ] {};

  \draw[dtarrow]
    (obs) -- (policy);

  \draw[dtarrow]
    (body) -- (reward);

  \draw[dtarrow]
    (reward.west) -- (update.east);

  \draw[dtfeedback]
    (update) -- (policy);

  \draw[dtfeedback]
    (body.south) -- ++(0,-9mm) -|
    node[pos=.52, below]{sensory consequence}
    (obs.south);

\end{tikzpicture}
\end{adjustbox}

\caption[Reinforcement learning coupled to embodied neural control.]
{Proposed division between fast closed-loop sensorimotor dynamics and
slower reinforcement learning. The learned policy can modulate neural
inputs, coupling strengths, sensory gains, or motor transformations.
Observed consequences and rewards drive policy or parameter updates
across repeated interactions.}
\label{fig:rl-adaptation}
\end{figure*}

The relationship between fast sensorimotor dynamics and slower adaptive
learning is summarized in Figure~\ref{fig:rl-adaptation}. The policy can
act through the neural controller rather than bypassing it, allowing RL
to tune interpretable controller parameters while physical behavior
continues to emerge from the closed neural--body--environment loop.

RL has already been applied to muscle-driven neuromechanical simulations
to generate locomotor behaviors and investigate principles of motor
control \cite{song2021deep}. Nevertheless, successful task performance
does not by itself establish biological validity. Learned movement can
depend strongly on the choice of reward function, anatomical model,
muscle parameters, training distribution, and regularization.
Consequently, policies should be evaluated not only by task performance
but also by comparisons with experimental kinematics, kinetics, muscle
activity, sensory responses, and energetic expenditure.

We envision RL serving not merely as a control algorithm but as a bridge
between neural dynamics and adaptive behavior. Future digital twins
could combine biologically inspired neural models, such as coupled
FitzHugh--Nagumo systems, with learning mechanisms that modify coupling
strengths, sensory gains, controller parameters, or motor-output
transformations based on experience. Such a hybrid architecture could
retain interpretable neural dynamics while allowing adaptation at a
slower learning timescale. It could therefore provide a computational
framework for investigating motor learning, behavioral adaptation,
rehabilitation, and skill acquisition. This proposed combination remains
a research direction, however, rather than an established biological
model of synaptic learning.

Reinforcement learning may also contribute to the integration of digital
twins with robotic experimentation. Policies initially trained in
simulation could be refined using observations obtained from physical
robots or biological experiments. However, differences between simulated
and physical systems create a ``reality gap.'' Approaches such as system
identification, domain randomization, uncertainty-aware training, and
sim-to-real adaptation may be required before policies learned in a
digital twin can be safely applied to a physical platform
\cite{kober2013reinforcement,tobin2017domain}. In experiments involving
animals, policy exploration must additionally be constrained by
experimental safety, animal-welfare requirements, and human oversight.

More broadly, machine-learning methods may support continuous estimation
of neural, sensory, and musculoskeletal parameters, while robotic
platforms can provide standardized data acquisition and empirical model
validation. The resulting workflow could alternate among observation,
model calibration, policy optimization, experimental testing, and model
revision. Parameter adaptation should include identifiability analysis
and uncertainty quantification so that changes in model parameters are
not mistakenly interpreted as uniquely determined biological mechanisms
\cite{villaverde2019benchmarking}.

Looking further ahead, neuro-musculoskeletal digital twins may evolve
from passive simulation environments into active scientific instruments.
Robot-scientist systems have demonstrated that automated platforms can,
within restricted experimental domains, generate or select hypotheses,
execute experiments, analyze results, and use those results to guide
subsequent experiments
\cite{king2009automation,sparkes2010robot}. Coupling these capabilities
to a neuro-musculoskeletal digital twin could support model-based
selection of informative perturbations, prediction of behavioral
responses, and iterative refinement of mechanistic hypotheses.

Predictions concerning disease progression, interventions, or therapeutic
outcomes would require disease-specific data, prospective validation, and
explicit representation of predictive uncertainty
\cite{saxby2023digital}. Thus, the near-term role of such twins is more
appropriately framed as prioritizing hypotheses and experiments rather
than replacing physical experimentation. With continued integration of
multimodal biological data, adaptive learning, biomechanics, and robotic
experimentation, neuro-musculoskeletal digital twins could nevertheless
provide an important foundation for computational neuroscience and
embodied artificial intelligence.

\section*{Conclusion}

We have argued that future animal digital twins should be viewed as
embodied dynamical systems integrating neural activity, musculoskeletal
mechanics, sensory feedback, and environmental interactions within a
unified computational framework. Rather than treating these components
as independent subsystems, we propose that behavior emerges from their
continuous and reciprocal interactions through closed-loop sensorimotor
dynamics
\cite{chiel1997brain,beer2000dynamical,ijspeert2023integration,
ramdya2023neuromechanics}.

To support this vision, we introduced a conceptual framework for a
neuro-musculoskeletal digital twin of the mouse that combines multimodal
anatomical reconstruction, biomechanical simulation, tactile sensing,
and Bonhoeffer--van der Pol/FitzHugh--Nagumo neural dynamics. Existing
whole-body mouse musculoskeletal models provide an important foundation
for such closed-loop neuromechanical simulations
\cite{ramalingasetty2021mouse}. The BVP/FitzHugh--Nagumo model provides
a reduced dynamical-systems representation of excitation, recovery,
refractoriness, and oscillatory behavior while remaining computationally
tractable
\cite{fitzhugh1961impulses,nagumo1962active}. It may therefore provide
a useful intermediate level of abstraction between detailed
conductance-based neural models and purely engineering-oriented
controllers.

Within this framework, locomotion, tactile exploration, motor
coordination, and behavioral adaptation are interpreted as emergent
phenomena arising from interactions among neural dynamics, body
mechanics, sensory transduction, and environmental contacts. This
interpretation does not imply that a single oscillator model can capture
all relevant neural processes. Rather, the BVP/FitzHugh--Nagumo model
serves as an initial reduced-order representation that can subsequently
be extended or replaced when specific questions require detailed
cellular, synaptic, or network mechanisms.

A key aspect of the proposed framework is the integration of multimodal
imaging data, including X-ray CT, high-resolution white-light sectioning,
and Scx-GFP imaging. CT provides skeletal geometry, white-light sections
provide complementary soft-tissue anatomy, and Scx-GFP reporters can
help identify tendon-associated structures
\cite{pryce2007transgenic}. Following image registration, segmentation,
and anatomical annotation, these complementary datasets could provide
the basis for constructing subject-specific models of skeletal,
muscular, and connective-tissue structures. Such models may enable
increasingly anatomically grounded simulations of movement, force
transmission, sensation, and sensorimotor control. Their accuracy,
however, will depend on quantitative validation of tissue segmentation,
muscle paths, attachment sites, mechanical parameters, and predicted
behavior.

More broadly, we envision neuro-musculoskeletal digital twins as a
convergence point for computational neuroscience, biomechanics,
artificial intelligence, and robotics. Adaptive learning methods may
support the acquisition of sensorimotor policies and continuous model
calibration
\cite{song2021deep,villaverde2019benchmarking}, whereas robotic
experimentation could provide standardized behavioral measurements and
iterative empirical validation. Autonomous scientific platforms have
already demonstrated closed-loop cycles of hypothesis selection,
experiment execution, data interpretation, and model refinement in
restricted experimental domains
\cite{king2009automation,sparkes2010robot}. Extending these principles
to behavioral neuroscience remains a long-term objective requiring
robust uncertainty quantification, prospective validation, animal-welfare
safeguards, and human oversight.

Future digital twins could continuously assimilate experimental
observations, update model parameters, compare alternative hypotheses,
predict responses to candidate perturbations, and identify informative
subsequent experiments. Predictions of disease progression,
intervention effects, or therapeutic outcomes would additionally require
disease-specific data and prospective biological validation
\cite{saxby2023digital}. Accordingly, the near-term role of these systems
should be understood as supporting hypothesis prioritization and
experimental design rather than replacing physical experiments.

Ultimately, neuro-musculoskeletal digital twins represent more than an
extension of conventional biomechanical simulation. They offer a
framework for investigating how biological intelligence emerges through
the coupling of nervous systems, physical bodies, sensory processes, and
environments. By unifying anatomy, dynamics, sensation, learning, and
behavior within a common computational architecture, such systems may
advance the study of embodied intelligence, accelerate model-guided
scientific discovery, and inform the development of adaptive robotic and
biomedical technologies.

\clearpage

\begin{acknowledgements}
The authors thank their collaborators and colleagues for valuable
discussions on mouse anatomy, multimodal imaging, musculoskeletal
modeling, computational neuroscience, and robotics. The authors also
acknowledge the researchers and developers who have made anatomical
datasets, biomechanical models, and open-source simulation platforms
available to the scientific community.

This work was supported by the RIKEN Challenge Grant. The funder had
no role in the conceptualization or preparation of the manuscript or
in the decision to publish it.
\end{acknowledgements}

\bibliographystyle{zHenriquesLab-StyleBib}
\bibliography{library}

\onecolumn
\newpage




\end{document}